# Tracking and Predicting Evolution of Social Communities

Mark Goldberg, Malik Magdon-Ismail, Srinivas Nambirajan, James Thompson
Computer Science Deptartment, Renssalear Polytechnic Institute, Troy, NY 12180.
Email: {goldberg, magdon, srinivas,thompja}@cs.rpi.edu

***Abstract*—We develop an algorithmic framework for studying the evolution of communities in social networks. We begin with the theoretical foundation, from which we conclude that an evolution is at most as strong as its weakest link. This allows us to formulate an efficient algorithm to identify *all* evolutionary sequences in a dynamic social network. We use this algorithm to empirically study community evolution in several large social networks, to identify those features of the early stages of a community that indicate whether a community is going to be short-lived or not. Our results show that it is possible to correlate the lifespan of a community to *structural* parameters of its early evolution; these conclusions are robust across all the social networks we have investigated.**

## I. Introduction

Large data repositories have opened new possibilities in social network analysis. It is necessary to develop efficient and accurate algorithms that can bring important features of a network to the forefront. Communities are fundamental units of every social network; their structure and evolution are essential to understanding the structure and functionality of large networks.

To study community evolution, one must first detect the communities; detecting communities should be based on a definition of what groups of users qualify to be called a community. Detecting all the communities in a large social network remains unsolved, due to challenges ranging from the computational burden (some formulations require the solution of NP-hard optimization problems [15]) to the conceptual definition of communities (a definition of a social community should emphasize that communities are locally defined and can overlap [2, 8], which may yield an excessive number of communities). Many algorithms have been proposed that attempt to discover communities, for example [2, 6, 7, 9, 15, 17]. We empirically study the effect of several such algorithms on our understanding of community evolution. Our contributions are:

(i) A foundational framework for studying community evolution, *assuming* that all communities at given instances of time have been provided by some algorithm. We propose three basic axioms for the definition of an evolution which imply the consecutive link approach to detecting evolutions; in particular, the axioms imply that *an evolution is at most as strong as its weakest link*. We use this thesis to develop a low order, polynomial time, dynamic programming algorithm for identifying *all* evolutions in a network based on the construction of a multi-partite *evolution graph*. Given a community, we can identify its entire evolution tree, *i.e.*, the set of valid evolutions for the community.

(ii) Empirical evaluation. We use several well-known algorithms for community detection to analyze four large social networks: DBLP (a social network of collaborations among CS researchers); IMDB (a social network of collaborations among actors); BLOGS (a discussion social network from the LiveJournal blogosphere); WIKI (a social-semantic collective wisdom network).

(iii) A framework for prediction. Using our detected community evolutions, we focus on the lifespan of the communities as one interesting variable to predict. In particular, we ask what properties of the early evolution of a community are indicative of the ultimate lifespan of the community. We identify that the *size* and *intensity* of a community are among the two most predictive properties. Our results are performed within a leave-one-out cross validation framework to ensure that overfitting of the data is minimized, and hence the results are statistically reliable.

*Prior Work.* Evolution of communities has recently become active. [4, 18] formulate a community as a hidden variable, and the communications (or interactions) between agents as observed. [3, 11] formulate community detection as an optimization of a weighted sum of cost functions measuring the evolutionary continuity and the community quality. [14] uses Laplacian dynamics to quantify community cohesiveness. [10] defines "natural" communities as groups of nodes whose cohesiveness is largely unaffected by small permutations in a network.

An (exponential) algorithmic formulation was given in [19], which proposes that a community evolution can be constructed by defining a community over time. To identify and characterize community evolutions, [1] utilizes a set of behavioral evolution events.

Our work is algorithmic in nature, based on an axiomatic foundation for the evolution of communities.

Our algorithms are efficient (low-order polynomial) and hence can be applied to very large networks. Our work focuses on tracking the evolution when given the communities. To actually obtain the communities, we use existing algorithms as mentioned previously. [16] takes an algorithmic approach toward quantifying social group evolution specifically for the clique percolation community detection algorithm. It is not clear how that framework can apply to evolution in general, since it relies on specific properties of *how the communities are constructed*. Our philosophy is that tracking of evolutions should only need the communities.

## II. Axiomatic Foundations for Evolution

**Notation.** We consider dynamic communities in social networks. Time is discretized into time steps $t = 1, \ldots, T$. Each time step has a physical duration, $\tau$ (for example $\tau = 1$ week in blogs); $\tau$ should be chosen to match the time scale of community dynamics that one is interested in. At every time step $t$, let $\mathcal{C}_t$ denote the set of communities observed, $\mathcal{C}_t = \{C_{1,t}, \ldots, C_{n_t,t}\}$ ($n_t$ is the number of communities at step $t$). Given a community $X_0 \in \mathcal{C}_t$, an evolution of $X_0$, or *chain* $P(X_0)$, is a sequence of communities $X_0, X_1, \ldots, X_k$ with $X_i \in \mathcal{C}_{t+i}$. The length of the chain is $k$; the parent is $X_0$ and the leaf of the chain is $X_k$.

Our goal is to discover all chains which are valid evolutions of $X_0$. Formally, assume there is an oracle $F(\cdot)$ which takes as input a chain $P(X_0)$ and outputs its strength $F(P) \in [0, 1]$. The strength $F(P)$ measures how plausible it is that the chain $P$ is as an evolution of community $X_0$. Then, the chain is accepted as a *valid* evolution if its strength is above a (user defined) threshold $\lambda$. Generally, we are interested in the *maximal* valid chains. A valid chain is maximal if it is not a proper subchain of any other valid chain.

**Problem Statement.** Given the sets of communities at each time step, $\mathcal{C}_1, \mathcal{C}_2, \ldots$, determine all the valid maximal chains. The figure illustrates the general setup.

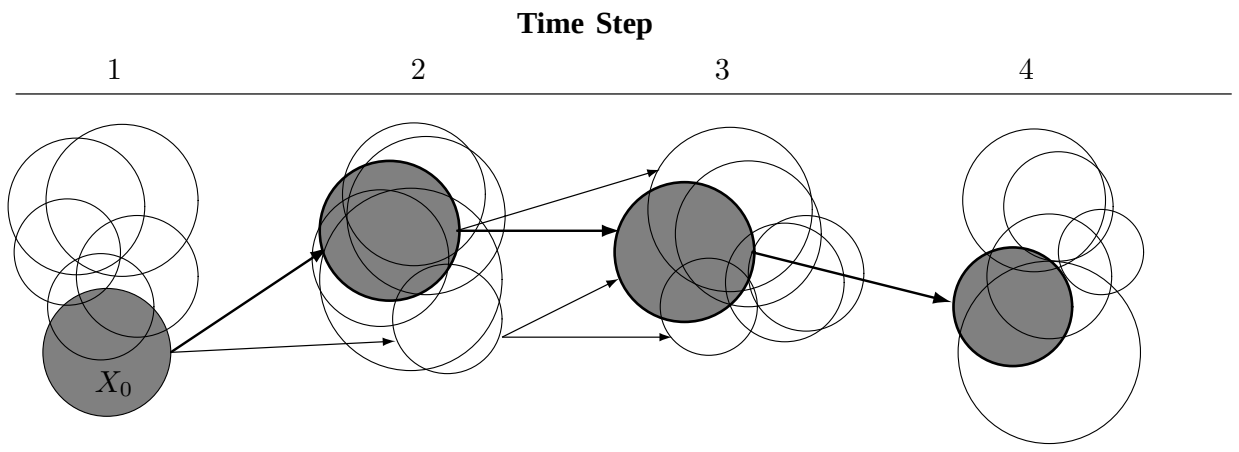

We have communities at each time step. Illustrated is one particular community $X_0$, and all the valid maximal chains for which $X_0$ is the parent. The shaded chain is the longest valid chain for which $X_0$ is the parent. Our experimental study will focus on these longest chains, though we would like to develop algorithms to find all valid chains. Even given the strength oracle $F$, this is a non-trivial problem. The brute force approach would test every chain for validity; since there might be exponentially many chains, this is not computationally feasible. To develop a useful algorithm, we first study the properties of the chain strength oracle $F(\cdot)$.

We seek the simplest self evident properties that $F(\cdot)$ should have. Part of our contribution is to show that these simple properties alone are enough to derive non-trivial implications that allow us to identify evolution in real data. We will show that, given some intuitive axioms that the oracle $F(\cdot)$ should satisfy, there is a simple characterization of $F$. This characterization reduces our problem to that of detecting valid evolutions of length 1, from which we will develop an efficient dynamic programming algorithm to solve our full problem.

### A. Chain Strength Axioms

**Axiom 1 [Identity].** For any community $X$, any chain obtained by repeating $X$ has maximum strength:

$$F(X, X, \ldots, X) = 1.$$

**Axiom 2 [Monotonicity].** A chain's strength is at most the strength of any subchain. So, for $0 \le i < j \le k$,

$$F(X_0, X_1, \ldots, X_k) \le F(X_i, X_{i+1}, \ldots, X_j).$$

**Axiom 3 [Extension].** If two valid chains have a community in common, then the chain constructed by *extending* the prefix (up to the common community) of one chain with the suffix (starting at the common community) of the other chain is valid. Specifically, let $X_0, \ldots, X_k$ and $Y_0, \ldots, Y_\ell$ be two chains, with $X_i = Y_j$ for some $i \le k$ and $j < \ell$. Then, for every $\lambda > 0$,

$$\begin{matrix} F(X_0, \ldots, X_k) \ge \lambda \\ F(Y_0, \ldots, Y_\ell) \ge \lambda \end{matrix} \implies F(X_0, \ldots, X_i, Y_{j+1}, \ldots, Y_\ell) \ge \lambda.$$

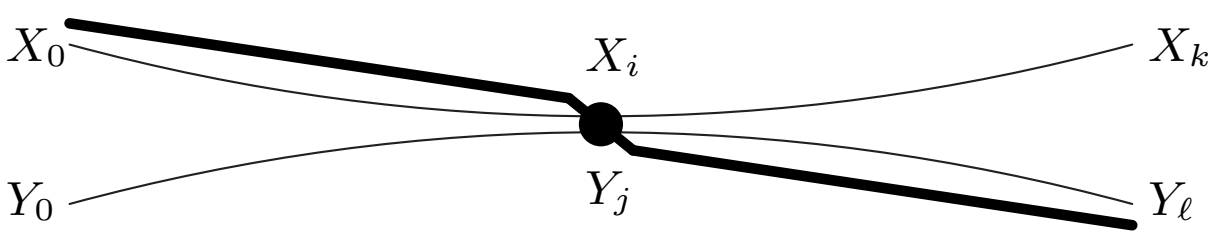

The intuition behind the third axiom is that the evolution of $X_0$ to $X_i$ is valid (because the whole $X$-evolution is valid); now, continuing this evolution along the $Y$-chain should be valid (because $X_i = Y_j$ and the whole $Y$-evolution is valid). These axioms are intuitive, and they imply a strong property about $F$.

**Theorem 1.** *If $F$ satisfies the identity, monotonicity and extension axioms, then, for any chain $X_0, \ldots, X_k$,*

$$F(X_0, \ldots, X_k) = \min_{i=0,\ldots,k-1} F(X_i, X_{i+1}).$$

We should note that our theorem refers to an abstract oracle $F$ which can take as input any chain $X_0, \ldots, X_k$

of arbitrary length $k$ and output its strength. Of course, in the realized sets of communities $\mathcal{C}_1, \mathcal{C}_2, \ldots$, not every possible chain is present. The oracle $F$ would only be applied to those chains that appear in the time sequence of communities. To paraphrase Theorem 1, *the strength of a chain is the strength of its weakest link*; an evolution is valid if and only if every step in the evolution is valid. An immediate consequence of Theorem 1 relates to constructing the valid evolutions. If the evolution $P = X_0, X_1, \ldots, X_k$ is valid and ends at time step $t$ (so $X_k \in \mathcal{C}_t$), we can extend this to a valid evolution ending at time step $t+1$ by considering all communities at time $t+1$. For every $Y$ such that $Y \in \mathcal{C}_{t+1}$ and $F(X_k, Y) \geq \lambda$, the chain $P' = X_0, X_1, \ldots, X_k, Y$ is a valid evolution ending at time step $t+1$. It is immediately clear that to compute the valid evolutions we need only know how to compute the strength of all possible 1-step evolutions; i.e. we only need to specify the oracle $F(X, X')$ for two arbitrary sets $X, X'$.

**Corollary 2.** *An evolution $X_0, \ldots, X_k$ is valid if and only if $F(X_i, X_{i+1}) \geq \lambda$ for $i = 0, \ldots, k-1$.*

For the remainder of this paper, to make the discussion concrete, we choose an intuitive set intersection based measure for the strength of an evolution,

$$F(X, X') = \frac{|X \cap X'|}{|X \cup X'|}.$$

To study the properties of long evolution chains, we first need to construct all the maximal valid chains from the observed time series of communities in the network. Recall that a valid chain is maximal if it is not a proper subchain of some other valid chain.

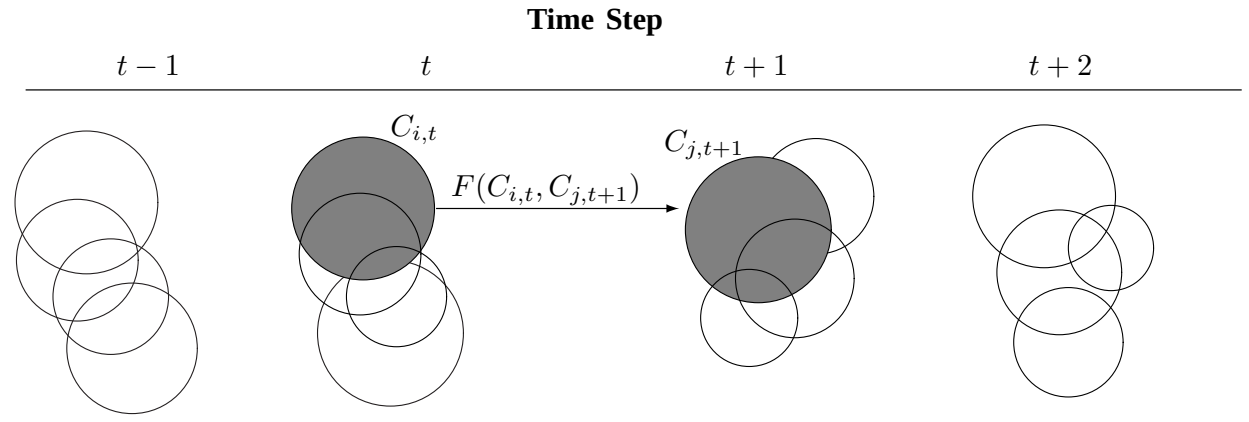

To compute all valid chains we first compute the weighted, multipartite graph $K = (\{C_{i,t}\}, E_K, F_K)$, illustrated above. The nodes are the communities in every time step. Edges $(C_{i,t}, C_{j,t+1})$ exist between communities in consecutive time steps that have non-empty intersections; the edge weight is $F(C_{i,t}, C_{j,t+1})$. Given $K$, the maximal paths with parent $C_{i,t}$ can be computed efficiently (See full version for details).

### B. Detecting Communities

If the communities are not given to us ahead of time, then we need to detect them. At time step $t$, if we have a set of interactions among the nodes, then we can construct communities using a graph clustering algorithm. Specifically, let $G_t = (V, E_t, W_t)$ be the social network graph at time step $t$; $V$, the set of nodes across all time steps; and $E_t$ is the set of interactions at time $t$ with weights in $W_t$. For example, consider the DBLP data. For a particular year (the time step) we two authors share an edge if they co-published a paper. The edge weight relates to the number of co-publications in that year, with more weight being given to co-publications that have fewer additional authors.

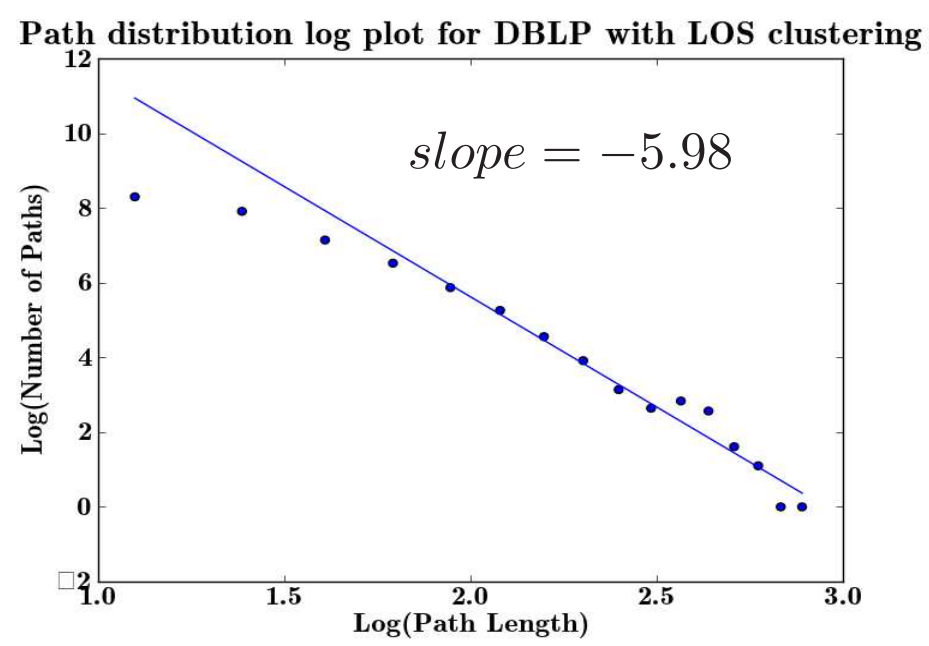

Fig. 1. Distribution of community lifetimes in DBLP using LOS clustering. The tail shows power law decay with exponent $\alpha = -5.98$

The graph $G_t$ is the input to a community detection algorithm. The output is $\mathcal{C}_t$, the set of communities at time step $t$. For social networks, any community detection algorithm should allow communities to overlap. We tested several community detection algorithms in our experiments: LOS[2], $K$-CP[17], RRW[12], FOG[5], and SSDE[13].

## III. Prediction

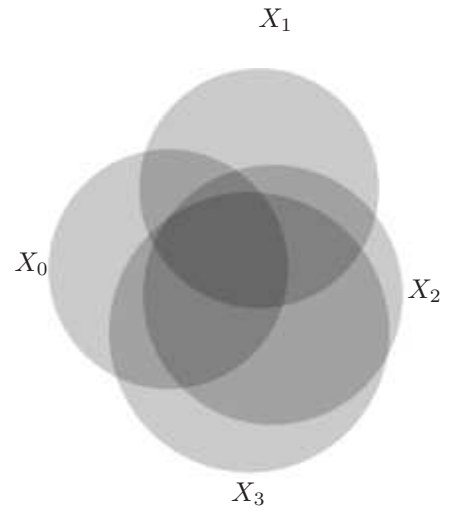

We consider the first four communities, $X_0 \ldots X_3$, of an evolution to predict the length of the evolution (see the figure). Let $s_0, s_1, s_2, s_3$ be the sizes of the communities, where $s_i = |X_i|$. We define the density of a community as the fraction of its authors' collaborative efforts spent within the community , $d_i = W_{in}(X_i)/(W_{in}(X_i) + W_{out}(X_i))$ (see appendix). Let $r_0, r_1, r_2$ be the sizes of the consecutive pairwise intersections, $r_i = |X_i \cap X_{i+1}|$. We define the cores $q_0, q_1$ as the intersection of these intersections, $q_i = |X_i \cap X_{i+1} \cap X_{i+2}|$. Finally, we define the hypercore as the intersection of the cores, $c_0 = |X_0 \cap X_1 \cap X_2 \cap X_3|$. These parameters relate to the shaded regions in the figure. From these parameters, we derive 26 features that characterize the early evolution of a community such as stability and rate of change of the community.

## IV. Experiments

We consider 4 datasets: coauthorship of academic papers (DBLP), movie co-stars data (IMDB), LiveJournal blog interactions (BLOG), and wikipedia edit data (WIKI). Each data set is a bipartite graph with users and objects. We construct communities over users in disjoint time intervals to find evolutions. The distribution of evolution lengths follows an inverse power law (Figure 1).

The choice of $\lambda$ influences the nature of detected evolutions. A high value of $\lambda$ divides true evolutions, while a low value of $\lambda$ classifies meaningless evolutions as valid ones. We determine $\lambda$ by comparing detected evolutions of real communities to those of random communities. We set $\lambda = 0.25$, at which point the majority of random evolutions have a lifespan of at most 3.

*Predicting Evolution* We consider predicting the lifespan of an evoution. We only consider maximal evolutions of length at least 4. The prediction task is to estimate, based on the features from section III, evolution length. We use a very simple linear-regression framework with leave-one-out cross validation (LOO-CV).

We may determine the features which are most useful for prediction by looking at how often it is found as a significant feature. We show the features which are found to be significant more than 40% of the time.

| Feature | Significance | Av. Weight |
|---|---|---|
| Density | 74% | 0.22 |
| Intersection | 57% | 0.29 |
| Size | 48% | -0.22 |
| Growth | 43% | -0.001 |
| Core | 43% | -0.11 |

## V. Conclusion

We have developed a framework for studying evolution of communities. Our results indicate that the evolutions of detected communities display powerlaw behavior which is not present in random communities, which means that the community detection algorithms detect non-random communities. In particular, these communities are more stable than random communities (as measured by average lifespan).

We then studied the predictability of evolution, in particular lifespan. We found a consistent set of features from the early evolution that can predict (out-sample) the lifespan of the community. These features were consistent over different data sets and community detection algorithms. In particular, density, intersection, and core size are quite significant and have strong positive correlation with lifespan; size shows strong negative correlation with lifespan. Conclusion: *intense, small, stable communities last longest*.

**Acknowledgements.** We thank R. Escriva (open source version of LOS), S. Kelly and K. Mertsalov (initial discussion and the BLOG data) and A. Lavoie (WIKI data). This material is based upon work partially sponsored by: U.S. DHS through ONR grant number N00014-07-1-0150 to Rutgers University and continues under the Army Research Laboratory under Cooperative Agreement Number W911NF-09-2-0053.

## References

[1] S. Asur, S. Parthasarathy, and D. Ucar. An event-based framework for characterizing the evolutionary behavior of interaction graphs. In *Proc. ACM SIGKDD*, 2007.

[2] J. Baumes, M. Goldberg, and M. Magdon-Ismail. Efficient identification of overlapping communities. In *ISI*, 2005.

[3] D. Chakrabarti, R. Kumar, and A. Tomkins. Evolutionary clusters. In *Proc. ACM SIGKDD*, 2006.

[4] H.-C. Chen, M. Goldberg, M. Magdon-Ismail, and A. Wallace. Reverse engineering an agent-based hidden markov model for complex social systems. *Int. J. Neu. Sys.*, 18(6):491–526, 2008.

[5] G. Davis and K. Carley. Clearing the fog: Fuzzy, overlapping groups for social networks. *Social Networks*, 30(3):201–212, 2008.

[6] S. Fortunato. Community detection in graphs. *Physics Reports*, 486(3-5):75 – 174, 2010.

[7] M. Girvan and M. Newman. Community structure in social and biological networks. *PNAS*, 99(12):7821–6, 2002.

[8] M. Goldberg, S. Kelley, M. Magdon-Ismail, and W. Wallace. Overlapping communities in social networks. In *Proc. 2nd SocialCom*, pages 103–113, 2010.

[9] S. Gregory. Finding overlapping communities using disjoint community detection algorithms. In *Complex Networks: CompleNet 2009*, pages 47–61, 2009.

[10] J. Hopcroft, O. Khan, B. Kulis, and B. Selman. Tracking evolving communities in large linked networks. *PNAS*, 101(1):5249–5253, 2004.

[11] Y.-R. Lin, Y. Chi, S. Zhu, H. Sundaram, and B. L. Tseng. Facetnet: A framework for analyzing communities and their evolutions in dynamic networks. In *WWW*, 2008.

[12] K. Macropol, T. Can, and A. K. Singh. Rrw: repeated random walks on genome-scale protein networks for local cluster discovery. *BMC Bioinformatics*, 10(283), 2009.

[13] M. Magdon-Ismail and J. Purnell. Ssde-clustering: Fast overlapping clustering of networks using sampled spectral distance embedding and GMMs. In *SocialCom*, 2011.

[14] P. J. Mucha, T. Richardson, K. Macon, M. A. Porter, and J.-P. Onnela. Community structure in time-dependant, multiscale, and multiplex networks. *Science*, 328, 2010.

[15] M. E. J. Newman. Modularity and community structure in networks. *PNAS*, 103(23):8577–8582, 2006.

[16] G. Palla, A.-L. Barabasi, and T. Vicsek. Quantifying social group evolution. *Nature*, 446(7136):664–667, 2007.

[17] G. Palla1, I. Derényi, I. Farkas, and T. Vicsek. Uncovering the overlapping community structure of complex networks in nature and society. *Nature, 814–818*, 2005.

[18] Y. Sun, J. Tang, J. Han, M. Gupta, and B. Zhao. Community evolution detection in dynamic heterogeneous information networks. In *Proc. KDD MLG*, 2010.

[19] C. Tantipathananandh, T. Berger-Wolf, and D. Kempe. A framework for community identification in dynamic social networks. In *Proc. 13th KDD*, 2007.